\documentclass[%
amsmath,amssymb,
superscriptaddress,
preprint,
showpacs,preprintnumbers,
 aps,
prb,
]{revtex4-1}

\usepackage[dvipdfmx]{graphicx}
\usepackage{epsfig}
\usepackage{dcolumn}
\usepackage{bm}
\usepackage{multirow}
\usepackage{amsmath}
\usepackage{longtable}
\usepackage{color}

\newcommand{\DFBO}{DyFe$_3$(BO$_3$)$_4$}

\begin{document}

\preprint{DyFe$_3$(BO$_3$)$_4$}

\title{Quadrupole moments in chiral material DyFe$_3$(BO$_3$)$_4$
observed by resonant x-ray diffraction}

\author{Hiroshi Nakajima}
\affiliation{
Division of Materials Physics, Graduate School of Engineering Science,
Osaka University, Toyonaka, Osaka 560-8531, Japan 
}
\affiliation{
Department of Materials Science, Graduate School of Engineering, Osaka
Prefecture University, Sakai, Osaka 599-8531 
} 

\author{Tomoyasu Usui}
\affiliation{
Division of Materials Physics, Graduate School of Engineering Science,
Osaka University, Toyonaka, Osaka 560-8531, Japan 
}

\author{Yves Joly}
\affiliation{Universit\'e Grenoble Alpes, Institut NEEL, F-38042 Grenoble, France}
\affiliation{CNRS, Institut NEEL, F-38042 Grenoble, France}

\author{Motohiro Suzuki}
\affiliation{
Japan Synchrotron Radiation Research Institute (JASRI), Sayo, Hyogo
679-5198, Japan
}

\author{Yusuke Wakabayashi}
\affiliation{
Division of Materials Physics, Graduate School of Engineering Science,
Osaka University, Toyonaka, Osaka 560-8531, Japan 
}

\author{Tsuyoshi Kimura}
\affiliation{
Division of Materials Physics, Graduate School of Engineering Science,
Osaka University, Toyonaka, Osaka 560-8531, Japan 
}

\author{Yoshikazu Tanaka}
\email{ytanaka@riken.jp}
\affiliation{
RIKEN SPring-8 Center, Sayo, Hyogo 679-5148, Japan}

\date{\today}

\begin{abstract}
By means of circularly polarized x-ray beam at Dy $L_3$
 and Fe $K$ absorption edges, the chiral structure of the electric 
 quadrupole was investigated for a single crystal of \DFBO\ 
in which both Dy and Fe ions are arranged in spiral manners.
The integrated intensity of the resonant x-ray diffraction of
 space-group forbidden reflections 004 
 and 005 is interpreted within the electric dipole transitions 
from Dy $2p_{\frac{3}{2}}$ to $5d$ and Fe $1s$ to $4p$, respectively.
We have confirmed that the handedness of the crystal observed at Dy
 $L_3$ and Fe $K$ edges is consistent with that observed at Dy $M_5$
 edge in the  previous study.  
By analyzing the azimuth scans of the diffracted intensity, 
the electronic quadrupole moments of Dy $5d$ and Fe $4p$ are derived.
The temperature profiles of the integrated intensity of 004 at the Dy
 $L_3$ and the Fe $K$ edges are similar to those of Dy-O and Fe-O bond
 lengths, while that at the Dy $M_5$ edge does not. 
The  results indicate that the helix chiral orientations of quadrupole
 moments due to Dy $5d$ and Fe $4p$ electrons are more strongly affected 
 by the crystal fields than Dy $4f$.
 \end{abstract}


\maketitle

\section {Introduction}

Chirality is one of the most important concepts over a wide range of
science, including particle physics, cosmology, biology, pharmacy,
condensed matter physics, and industry,
etc.~\cite{Wagniere2008cua,Barron2008cl} 
The key issue of chirality, or handedness, is the break of symmetry that 
plays a crucial role in a variety of the fields.
In condensed matter physics, the break of symmetry often gives an
excellent arena to manipulate the physical properties.  One prominent
example is found in the magnetoelectric multiferroic materials, where 
the ferroelectricity appears as a result of the phase transition
inducing such a magnetic order that breaks the inversion
symmetry.~\cite{Kimura2007smm,Cheong2007md}  
Many of them have a cycloidal or a screw spin structure, of which
handedness determines the sign 
of the spontaneous electric polarization.
In such a system, the multipole moments demonstrate the sign
of handedness together with the atomic and spin structure.
The electric multipole is represented by the density of the electron
cloud at an ion which is expanded with the spherical tensor $T^K_Q$.
Here $K$ represents the rank of tensor, and $Q$, $-K \leq Q \leq K$,
represents its projection. 
In principle, the electric multipole $\left < T^K_Q \right >$ is visible
with an x-ray beam because the electron has the large cross-section for
the x-ray beam.
In reality, however, observing a motif of the multipoles in crystal is
not easy in x-ray diffraction. 
In the conventional x-ray diffraction, which is used to observe the
crystal structure, the diffraction profile is mainly emerged from the 
core electrons of the consisting elements in the matter, 
where  the contribution of the valence electrons is negligible.
Whereas, when the system has the antiferroic type of the multipole
order, where the symmetry of valence electrons breaks the crystal
symmetry, the order is visible with the high intense x-ray beam from the 
synchrotron radiation source. 
For example, it has been found that the antiferroic quadrupole order in
CeB$_6$ at low temperatures shows  
tiny superlattice reflections $\left ( \frac{h}{2},\frac{k}{2},\frac{l}{2} \right )$,
where $h$, $k$, and $l$ are  odd numbers.~\cite{Yakhou2001k,Tanaka2004daq}

Since the discovery of linear dichroism~\cite{Templeton1980pxa} and
forbidden reflections~\cite{Templeton1986xba} at an
absorption edge, resonant x-ray diffraction (RXD) has been developed to
investigate a variety of ordered states, such as magnetic, charge, or
orbital orders. 
The atomic picture of the resonance is well known: an incoming photon
promotes a core electron to empty states, and it returns to
the same core hole, emitting a second photon of the same energy as the
incoming one.  The scattering length at an absorption edge, which is
sensitive to the polarization in the primary and secondary x rays,
carries site-specific information on unoccupied valence states on and
around the resonant ion. This sensitivity provides another useful aspect 
of RXD as described below.  

Circular dichroism in RXD has been found for
 low-quartz,~\cite{Tanaka2008rho,Lovesey2008rdc} which has  an
 enantiomorphic space-group pair, $P3_121$ (right-handed screw) and
 $P3_221$ (left-handed screw). 
 In crystals having the space-group pair $P3_121$  and $P3_221$,
 reflections $00l$ ($l \neq 3n$, $n = $integer) 
 are forbidden in non-resonant x-ray diffraction, but allowed in RXD for 
 certain x-ray energies 
 because of the sensitivity of the atomic scattering length to the x-ray
 polarization. 
 By carefully examining the space-group forbidden reflections observed
 with circularly-polarized x rays, 
 one can identify the absolute sign of the crystallographic helix
 chirality.~\cite{Tanaka2010doa,Tanaka2010dsc} 
For the ferromagnetic or the ferroelectric states, we hardly know the
 state of the valence electrons by x-ray diffraction,
because the symmetry of the magnetic order or the charge order is the
 same as the crystal symmetry.   
However, for the system having the screw-axis or the glide-plane, 
we can scrutinize the ordering state of the multipole moment, observing
 the space-group forbidden reflections,~\cite{Dmitrienko1983frd} even
 though the symmetry is the same as the crystal symmetry.

Recently we have successfully shown that RXD using circularly polarized
x rays not only identifies the absolute sign of the crystallographic
helix chirality, but also provides direct information on the multipole
moments accompanied with the crystallographic helix
chirality.~\cite{Usui2014ooq} 
We have determined two components among five of the $4f$ quadrupole
moment of Dy in \DFBO\ where the motif of the $4f$ quadrupole moment
coincides with the crystallographic helix chirality. 
In the experiment, we have observed forbidden reflection 001 with
circularly polarized soft x-ray beam at Dy $M_5$ absorption edge, where
the x rays enhance the resonance from the $3d_\frac{5}{2}$ state to the
vacant $4f$ state.  
 Accordingly the signal observed in the diffraction
 gives the information of the $4f$ quadrupole moment of Dy ions.

In the present study, we demonstrate that circularly polarized hard
x-ray beam at two absorption edges,  Dy $L_3$ and  Fe $K$, is also
useful to determine the absolute sign of crystal chirality in \DFBO\ as 
the soft x-ray beam at  Dy $M_5$ absorption edge.
The resonance from Dy $2p_\frac{3}{2}$ to  $5d$ at Dy $L_3$ absorption
edge, and  that from Fe $1s$ to $4p$ at Fe $K$ absorption edge give the
information of Dy $5d$  
quadrupole moment and the Fe $4p$ quadrupole moment, respectively.
We discuss the possibility of the birefringence phenomenon and the
higher-order transition like $E1E2$ for the observed data.
We also discuss the chirality in terms of the electric Dy $5d$
quadrupole moment and the Fe $4p$ quadrupole moment together with the
deformation of the crystal structure  as a function of temperature. 

We give a general background of resonant x-ray diffraction in
Sec.~\ref{sec:resonant}\@. 
We describe the crystal structure of \DFBO\
as well as the physical properties in Sec.~\ref{sec:structure} and 
the experimental geometry in Sec~\ref{sec:experiment}\@.
In Sec.~\ref{sec:result}, we show the experimental results of the
resonant diffraction both at Dy $L_3$ and Fe $K$ absorption edges: x-ray
absorption spectra, azimuth angle scans, and the temperature dependence
of the integrated intensity of forbidden reflection 004.
In Sec.~\ref{sec:discussion}, we analyze the azimuth scan data and
derive the Dy $5d$ quadrupole moment and the Fe $4p$ quadrupole moment
with theoretical interpretations.  In addition, we discuss the
temperature dependence in terms of the deformation of DyO$_6$ trigonal
prism and FeO$_6$ octahedron. 
 
\section{Resonant X-ray Diffraction}\label{sec:resonant}
The determination of the absolute structure of enantiomers, which have
exactly the mirror image to each other, is not easy because they have
the same chemical formula and atom to atom arrangements,  
and hence has been  considered as an important challenge of
crystallography.
Among many methods applied for the determination so far, 
x-ray diffraction with dispersion corrections has played an important 
role for a long time. 

The scattering length of x-rays for an atom can be written as
\begin{equation}
f = f_0 + f' + if '',
\label{equ:dispersion}
\end{equation}
where $f_0$ is the energy-independent x-ray scattering length and
corresponds to the Fourier transform of the electron density around the 
atom, $f'$ and $f''$ are the real and the imaginary parts, respectively,   
of the dispersion correction.
Note that the sign of $f''$, which is positive in the conventional x-ray
diffraction,  is negative in our convention with the phase factor
$e^{i(\mathbf{K} \cdot \mathbf{r}-\omega t)}$
with the momentum transfer $K$ from the x rays to the crystal. 
Since the absolute configuration of the enantiomorphic compounds was
firstly determined for tartaric acid~\cite{Bijvoet1951doa} using the
dispersion correction, this method has developed after the advent
of the synchrotron radiation source and is known as Multi-wavelength
Anomalous Diffraction (MAD),~\cite{Karle1980sdi, Hendrickson1985} 
and is  extensively developed to study the absolute configurations for
biochemical compounds.  The method requires to measure the diffraction
pattern at several energies around the $K$ or $L$ absorption edges of
the resonant scattering elements. 

Our method is quite different from the above.  We observe
\emph{space-group forbidden} reflections at the resonant energy in the
vicinity of the absorption edge, where the scattering length  is
sensitive to the polarization in the primary and secondary x rays. 
Some \emph{forbidden} Bragg indices turn to be \emph{allowed},  
when the equivalency of the local atomic configuration around the
resonant scatterer in a unit cell is broken by the screw axis  or the 
glide plane. 

The dispersion correction terms in the vicinity of the absorption edge 
are written as 

\begin{equation}
  f'+if''= m 
\sum_{n,g} \left(\frac{ \Delta}{\hbar}\right)^2 \frac{ 
   \left\langle g\left| \hat{o}_s^* \right| n \right\rangle \left\langle n \left | \hat{o}_i\right| g \right\rangle}
   {\hbar\omega - \Delta  + i\Gamma/2} \;,
\label{equ:resonance}
\end{equation}
\noindent where $g$ and $n$ label the ground and intermediate states,
respectively. The difference energy    
$\Delta=\mathcal{E}_n-\mathcal{E}_g$,  where $\mathcal{E}_g$ and
$\mathcal{E}_n$ are their corresponding energies, and $\left|g\right>$
and $\left|n\right>$ are their wave functions. 
The electron mass is denoted by $m$, $\hbar\omega$ is the photon
energy, $\Gamma$ is a phenomenological broadening depending on the
photoelectron kinetic energy and the core-hole width, and $i$ and $s$ label the incoming and
scattered waves. 
The transition operator, dominated by its electric part and taken to
second order, is given by 
\begin{eqnarray} 
 \hat{o} =  (\boldsymbol{\epsilon}\cdot\mathbf{r} )  \left \{
		     1+\frac{i}{2}
		      (\mathbf{k}\cdot\mathbf{r}) \right \}   \;,     
\label{eq:transion-operator}
\end{eqnarray}
The first term in the equation gives the electric dipole or $E1$
transition and the second one gives the electric quadrupole or $E2$
transition.  
 The compound \DFBO\ belongs to one of the
enantiomorphic space-group pair, $P3_121$ and $P3_221$ in the low
temperature phase, where the inversion symmetry is broken, hence the $E1E2$
term in Eq.~\ref{equ:resonance} is allowed for the forbidden reflections. 
However, the contribution is possibly negligible as is discussed later.

\section{Structure of D\lowercase{y}F\lowercase{e}$_3$(BO$_3$)$_4$} \label{sec:structure}
The compound \DFBO\ consists of DyO$_6$ trigonal prisms with a Dy$^{3+}$
ion in the center,  FeO$_6$ octahedra with a Fe$^{3+}$ ion in the center, 
and  BO$_{3}$ triangles with a B$^{3+}$ ion in the
center.~\cite{Hinatsu2003m} 
It belongs to trigonal space-group $R32$ (\#155) at room temperature.
The structure has an alternative stacking of BO$_3$ network planes and
Dy plus Fe planes along the $c$ axis. 
Space group $R32$ belongs to one of the 65 enantiomorphic space groups,
and includes the right-handed screws $3_1$ or  left-handed screws $3_2$
inside so that the crystal can have two choices for the atomic
configuration. 
Indeed, the chain  of the FeO$_6$ octahedra can form, in two ways, a
right-handed or a left-handed screw along the $c$ axis in $R32$.
These two atomic configurations, which are mirror images to each other, 
coexist in a single crystal with large chiral
domains.~\cite{Usui2014ooq} 
\begin{figure}[htp]
\includegraphics[width=80mm]{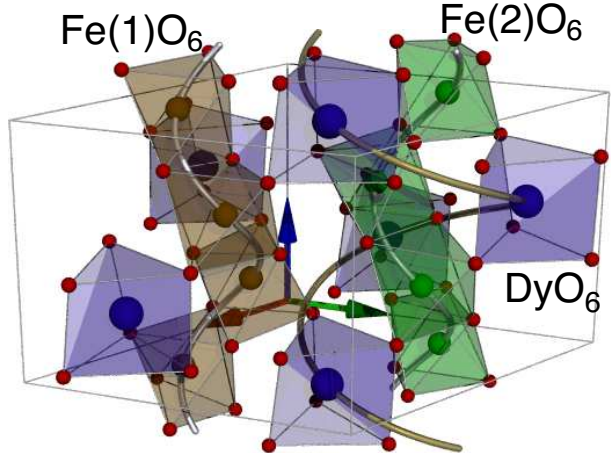}
\caption{
\label{fig:structure}
A view of the atomic configuration of  the \emph{left-handed} \DFBO\ in
 space group $P3_221$ (\#154).  
The  FeO$_6$ octahedra  and the  DyO$_6$ trigonal prisms are colored by
 brown and blue, respectively.  
 The Fe(1) ions are at the $3a$ site, and the Fe(2) are at the $6c$ site.
The unit-cell is shown by gray lines.
Yellow and gray helices represent spiral arrangements of Dy and Fe ions,
respectively.  
 }
\end{figure}
 
This compound has a first-order structural phase transition at 
$T_S \approx 285$  K,~\cite{Ritter2012m,Popova2008m} where  the
right-handed and left-handed structure in the space group $R32$ turn 
into the space groups $P3_121$ (\#152) and $P3_221$ (\#154),
respectively on decreasing  temperature. 
In these enantiomorphic space-group pair $P3_121$ and $P3_221$, the 
stacking chain of the Dy$^{3+}$ ions forms in a right-handed or a 
left-handed screw along the $c$ axis, respectively.  
Figure~\ref{fig:structure} displays the atomic structure of the 
\emph{left-handed} \DFBO\  in space group $P3_221$.
Note that the screw chain of the Dy$^{3+}$ ions in the low temperature
phase ($T< T_S$) succeeds the handedness of the chain of Fe$^{3+}$ ions
in the high temperature $R32$ phase ($T > T_S$).  
In the low temperatures, this compound has an additional phase
transition at $T_{\mathrm{N}} = 38$ K  
where both of Fe and Dy magnetic moments shows an antiferromagnetic (AF) 
order.~\cite{Ritter2012m} 
In the AF ordered phase, \DFBO\ exhibits a magnetoelectric effect
accompanied by a spin flop.~\cite{Popov2009o}

\section{Experimental}\label{sec:experiment}

The experiment was carried out at the beam line 29XUL at SPring-8 in
Harima, Japan.  A subsidiary experiment was carried out at the beam line
3A at PF in Tsukuba, Japan.  
A plate-like single crystal of \DFBO\ was
mounted on a copper holder in a liquid He flow-type cryostat on a
four-circle diffractometer.  The handedness of this crystal has been 
found to be \emph{left-handed}, space group $P3_221$ (\#154) in the
previous experiment.~\cite{Usui2014ooq}
The sample size was about 2 mm $\times$ 2 mm $\times$ 0.5 mm. The size
of the incident beam was 0.5 mm$\times$0.5 mm. 
A scintillation detector was used to observe the diffracted beam.  The
intensity of the incident beam was monitored by an ionization chamber.
The incident energy was tuned  by a Si (111)
double-crystal-monochromator followed by a pair of Si mirrors coated by
rhodium in order to cut the higher harmonic x rays at the beam line
29XUL. The helicity of the circularly polarized x-ray beam was
manipulated by a diamond phase retarder with 0.6 mm in thickness.  We
employed the $2\overline{2}0$ reflection of the diamond of which surface
is parallel to (111). In the experiment at PF, we used only a
$\boldsymbol{\sigma}$ linearly polarized x-ray beam. We measured the
intensity of space-group forbidden reflections 004 and 005 at two
absorption edges Dy $L_3$ at $E=7.796$ keV and Fe $K$ at $E=7.13$ keV in
a temperature range from $T=30$ K to $T=305$ K\@.  

\begin{figure}[htp]
\includegraphics[width=80mm]{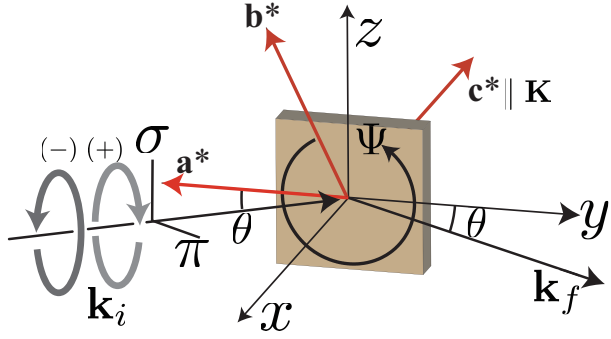}
\caption{
\label{fig:geometry}
(Color online)  A schematic view of the diffraction geometry with a
 right-handed  coordinates $x$, $y$,  and $z$.  The scattering vector 
 $\mathbf{K}=\mathbf{k}_{i}-\mathbf{k}_{f}$ is antiparallel to the $x$ 
 axis. The marks $(+)$ and $(-)$  denote the positive and negative
 helicity of the incident  beam, respectively. 
  Here $\mathbf{k}_{i}$  and $\mathbf{k}_{f}$ are the propagation
 vectors of the incident and diffracted x rays, respectively, and
 $\theta$  denotes the Bragg angle. 
The $\boldsymbol{\sigma}$   and $\boldsymbol{\pi}$ 
components are the unit vectors which represent the polarization of the
 incident beam.   
Here the $\boldsymbol{\sigma}$   component is perpendicular to the plane 
 of scattering, $\boldsymbol{\sigma}=(0, 0, 1)$, and the
 $\boldsymbol{\pi}$  component is parallel to the plane of scattering, 
 $\boldsymbol{\pi}=\left(\cos\theta,\,\sin\theta,\,0\right)$.  For the
 diffracted beam,  $\boldsymbol{\sigma}'=\boldsymbol{\sigma}$ and
 $\boldsymbol{\pi}'=\left(\cos\theta,\,-\sin\theta,\,0\right)$. 
}
\end{figure}

Our diffraction geometry is illustrated in Fig.~\ref{fig:geometry}.
The azimuth angle $\Psi$ is a rotation of the sample about the
scattering vector $\mathbf{K}=\mathbf{k}_{i}-\mathbf{k}_{f}$. The
vectors $\mathbf{k}_{i}$, and $\mathbf{k}_{f}$ are the propagation
vectors for the incident beam and the diffracted beam, respectively. 
We define the origin of the azimuth angle $\Psi=0$ with respect to the
direction of the reciprocal lattice vector 
$\mathbf{a}^{*}$ when it is parallel to the $-y$ axis, or 
$\mathbf{a}^{*}\Vert -\left(\mathbf{k}_{i}+\mathbf{k}_{f}\right)$,  
and the positive direction of $\Psi$ to be the counter-clockwise
rotation  as viewed looking up along the scattering vector
$\mathbf{K}$.  The polarization of the incident x-ray beam is defined by 
the unit vectors $\boldsymbol{\sigma}$ and $\boldsymbol{\pi}$. 

The average polarization state is expressed with the Stokes parameters
$P_{1}$, $P_{2}$, and $P_{3}$ where $P_{3}=+1$ and $P_{3}=-1$ correspond 
to the linear polarization parallel to the unit vectors
$\boldsymbol{\sigma}$ and $\boldsymbol{\pi},$ respectively, $P_{2}=+1$
and $P_{2}=-1$ correspond to the circular polarization,
represented by 
$\frac{1}{\sqrt{2}}\left(\boldsymbol{\sigma}+i\boldsymbol{\pi}\right)$
and
$-\frac{1}{\sqrt{2}}\left(\boldsymbol{\sigma}-i\boldsymbol{\pi}\right)$, 
respectively, $P_{1}=+1$ and $P_{1}=-1$ corresponds to the linear
polarization along the diagonal directions between 
$\boldsymbol{\sigma}$ and $\boldsymbol{\pi}$,
represented by 
$\frac{1}{\sqrt{2}}\left(\boldsymbol{\sigma}+\boldsymbol{\pi}\right)$
and
$\frac{1}{\sqrt{2}}\left(\boldsymbol{\sigma}-\boldsymbol{\pi}\right)$,
respectively. For the circularly polarized $P_{2}=+1$ ($-1$) state, the
$\boldsymbol{\pi}$ ($\boldsymbol{\sigma}$) component lags the
$\boldsymbol{\sigma}$ ($\boldsymbol{\pi}$) by $90^{\circ}$ according to 
the phase factor $\exp i\left(\mathbf{k}\cdot\mathbf{r}-\omega t\right)$. 
Here unit vectors $\boldsymbol{\sigma}$, $\boldsymbol{\pi}$, and 
$\hat{\mathbf{k}}_i=\mathbf{k}_i/|\mathbf{k}_i|$ satisfies the
right-handed rule $\boldsymbol{\sigma} \times \boldsymbol{\pi}
=\hat{\mathbf{k}}_i$. 
The polarization of the beam from the synchrotron radiation source is
usually well defined and that the relation 
$P_1^2 + P_2^2 + P_3^2 \cong 1$ holds. 

Resonant x-ray diffraction of space group forbidden reflections depends
on the geometry of the scattering system as well as the energy of the
x-ray beam. This is because it depends on the
polarization state of the x-ray beam as described by
Eq.~\ref{eq:transion-operator}, and the unit-cell structure factor is
described by a tensor of the atomic multipoles. 
 Hence the azimuth angle $\Psi$ scan gives important information not
 only about the symmetry of the local structure of the resonant ions,
 but also about the components of the atomic multipoles.

\section{Results}\label{sec:result}

\subsection{X-ray absorption spectra}\label{sec:absorption}

\begin{figure}[htp]
\includegraphics[width=80mm]{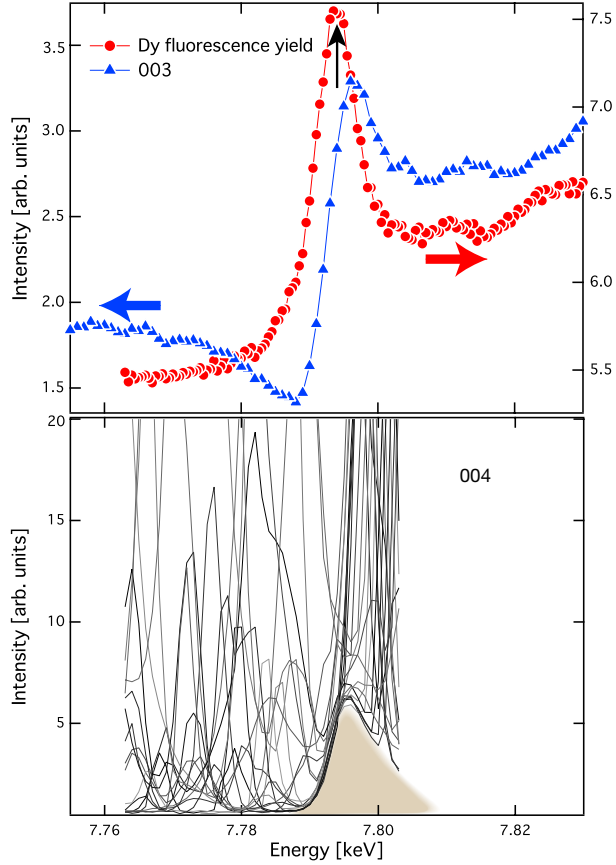}
\caption{
\label{fig:Dy-energy}
(Color online) Energy spectra of the x-ray absorption, the intensity of
reflection 003, and that of forbidden reflection 004 around the Dy
$L_3$ absorption edge. 
The data were observed at the beam line 3A in PF with the $\sigma$
polarized incident beam. 
Top: the x-ray absorption spectrum obtained in the total fluorescence 
mode is shown by red circles and the intensity of reflection 003 is
shown by blue triangles. The absorption effect is not corrected. 
The white line (the maximum absorption) is shown by a black arrow.
Bottom: the intensity of reflection 004 as a function of the x-ray
energy was observed for the azimuth angle $\Psi$ which was scanned in a  
range from 131.5 to 133.4$^{\circ}$ by every 0.1 degree step at $T=200$
K\@. 
The resonant effect underlying in  the intensity of reflection 004 is 
colored around $E=7.796$ keV. 
}
\end{figure}

\begin{figure}[htp]
\includegraphics[width=80mm]{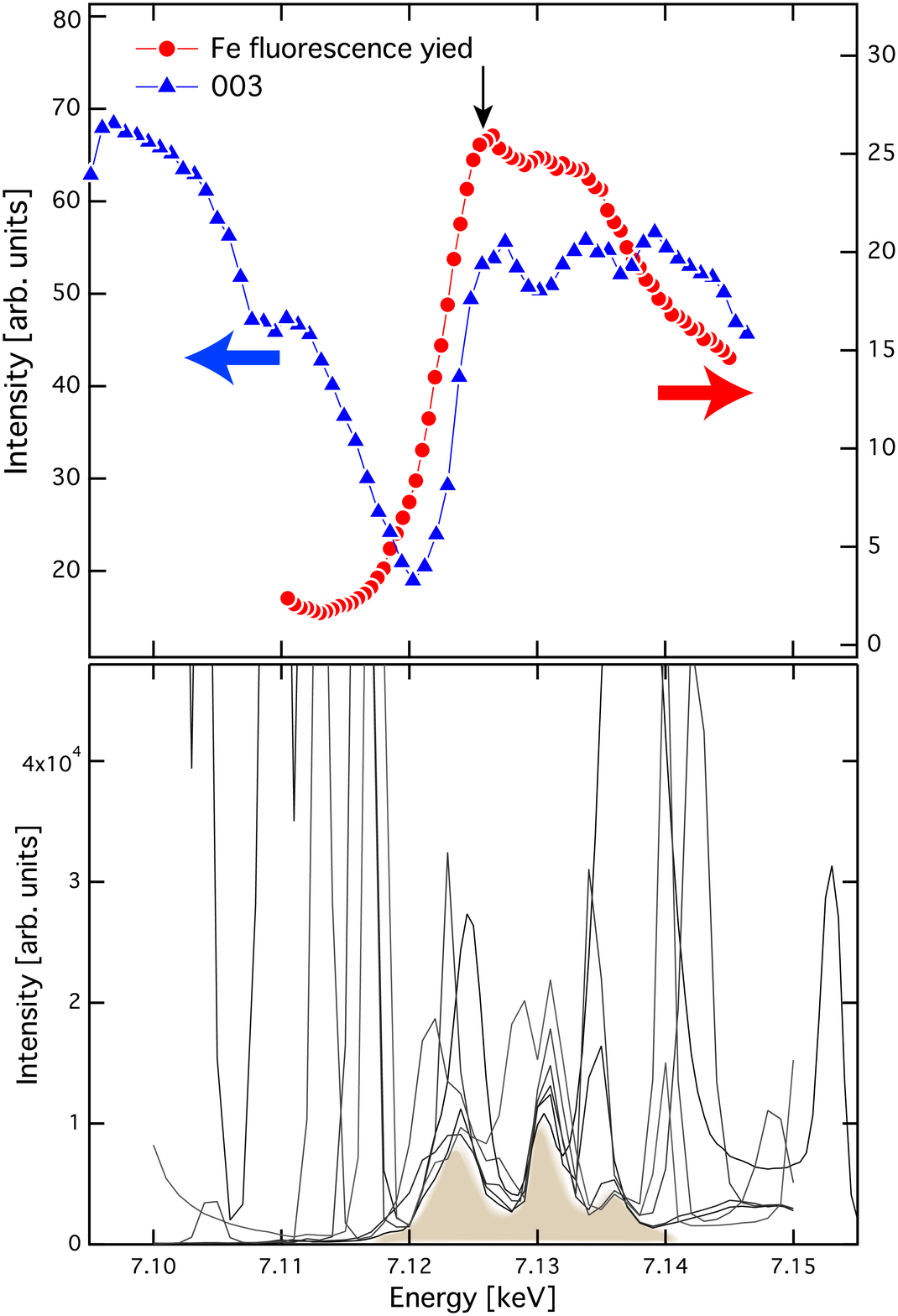}
\caption{
\label{fig:Fe-energy}
(Color online) Energy spectra of the x-ray absorption, reflection 003,
 and forbidden reflection 004 around the Fe $K$ absorption edge. 
The data were observed at the beam line 29XUL in SPring-8 with the $\pi$ 
polarized incident beam. 
Top: the x-ray absorption spectrum obtained in the total fluorescence
mode is shown by red circles and the intensity of reflection 003 is
shown by blue triangles. 
The absorption effect is not corrected. The white line (the maximum
absorption) is shown by a black arrow. 
Bottom: the intensity of reflection 004 as a function of the x-ray
energy was observed for azimuth angles $\Psi=-100,\, -105,\, -110,\,
-114,\, -115$, and $-126^{\circ}$ at $T=50$ K\@. 
The resonant effect underlying in  the intensity of reflection 004 is 
colored around $E=7.13$ keV.
}
\end{figure}

The x-ray absorption spectrum (XAS) and the energy dependence of the
space-group allowed reflection 003 in the vicinity of Dy $L_3$ and Fe
$K$ absorption edge are shown in top panels of Figs.~\ref{fig:Dy-energy}
and \ref{fig:Fe-energy}, respectively. 
The XAS were observed by measuring the total fluorescence yield.
The maximum absorption, so called the white line, of the Dy $L_3$ edge
and that of the Fe $K$ edge was found at $E=7.794$ keV and $E=7.125$
keV, respectively while the minimum intensity of reflection 003 around
the Dy $L_3$ edge and around the Fe $K$ edge was found at $E=7.788$ keV
and at $E=7.120$ keV, respectively.

We have successfully observed space-group forbidden
reflections 004 and 005 in the vicinity of both absorption edges.
The intensity of these forbidden reflections is about an order of
$10^{-4}$ to that of space-group allowed reflection 003.  
The resonant effect for the forbidden reflections is not easily specified 
from the x-ray absorption spectra and the energy spectrum of reflection
003 for both absorption edges.  
In order to observe the underlying resonant effect precisely,
we performed energy scans for forbidden reflection 004 at several 
azimuth angles $\Psi$ in a small range. 
Bottom panels of Figs.~\ref{fig:Dy-energy} and \ref{fig:Fe-energy}
show energy spectra of the intensity of reflection 004
around the Dy $L_3$ and the Fe $K$ edges, respectively.
As observed, reflection 004 is accompanied by strong multiple scattering  
effect or \emph{Umweganregung} effect which adds apparently random
bumpy structure to the spectra, which varies with the azimuth angle as
well as the x-ray energy.~\cite{Renninger1937,Weckert1997m} 
The underlying resonant enhancement is present under the intensity
curves of reflection 004, which are bumpy due to the multiple scattering
effect.  The same manner was taken for deducing the Te $L_1$ resonant
enhancement previously.~\cite{Tanaka2010doa} 
The enhancement for both absorption edges, observed as a single peak
around $E=7.796$ keV for the Dy $L_3$ edge and three peaks around
$E=7.13$ keV for the Fe $K$ edge, is shown by the colored areas in the
bottom panels of Figs.~\ref{fig:Dy-energy} and~\ref{fig:Fe-energy},
respectively.  The feature of three peaks around Fe $K$ absorption edge
has been reported in the literature, for example,
FeS$_2$.~\cite{Kokubun2004r} 
We employed $E=7.796$ keV for the Dy $L_3$ absorption edge and $E=7.13$ 
keV for the Fe $K$ absorption edge in the following experiments.  
 
\subsection{Azimuth angle scans}

We have performed azimuth angle scans for forbidden reflections 004 and 
005 at $E=7.796$ keV for the Dy $L_3$ absorption edge and $E=7.13$ keV 
for the Fe $K$ absorption edge with the positive and negative circular 
polarized x-ray beam. 
At each fixed azimuth angle $\Psi$, we measured the integrated intensity
by $\omega$ scan.  
The state of the x-ray polarization was manipulated by the diamond phase
retarder for both absorption edges; the Stokes parameters are summarized 
in Table~\ref{tab:polarization}.
The values are given from the deviation angle in $\theta$ for Bragg
reflection  $2\overline{2}0$ of the diamond retarder.
Here $(+)$ and $(-)$ represent the helicity positive ($P_2>0$) and the
helicity negative ($P_2<0$) states for the circularly polarized x-ray
beam, respectively.  Hereafter we use symbols $(+)$ and $(-)$ for the
positive and negative x-ray helicity, respectively. 
Because of an accidental inaccuracy for positioning the diamond phase
retarder, we found that the $P_3$, the linear polarization, had
unintended non-negligible values for both the positive and the negative
circularly polarized x rays, while the $P_1$, the diagonal linear
polarization, had negligible values. 
Accordingly, the values of $P_2$ are not exactly the same for both $(+)$
and $(-)$ states, except the sign. This small deviation in $P_2$,
however, does not matter for the following analysis for the quadrupole
moment as we discuss later.  

\begin{table}
\centering{}
\caption{
 The values of experimental Stokes parameters for two circularly
 polarized states, $(+)$ and $(-)$, at Dy $L_3$ ($E=7.796$ keV) and Fe
 $K$ ($E=7.13$ keV) absorption edges, together with the angle of the 
 diamond crystal, $\theta_{2\overline{2}0}$ for each polarization state. 
}
\begin{tabular}{c|c|cccc}
	\multicolumn{2}{c|}{} & $\theta_{2\overline{2}0}$ (degree) & $P_1$ & $P_2$ & $P_3$ \\
\hline
\multirow{2}{*}{Dy $L_3$ 7.796 keV} & $(+)$ & 39.111 & $-0.0152$ & $+0.978$ & $-0.114$ \\ \cline{2-6}
                                    & $(-)$ & 39.073 & $+0.0139$ & $-0.967$ & $-0.203$ \\ 
\hline
\multirow{2}{*}{Fe $K$ 7.130 keV}   & $(+)$ & 43.612 & $-0.0195$ & $+0.990$ & $-0.023$ \\ \cline{2-6}
                                    & $(-)$ & 43.561 & $+0.0182$ & $-0.986$ & $-0.110$ \\ 
\hline
\end{tabular}
\label{tab:polarization}
\end{table}

\begin{figure}[htp]
\includegraphics[width=80mm]{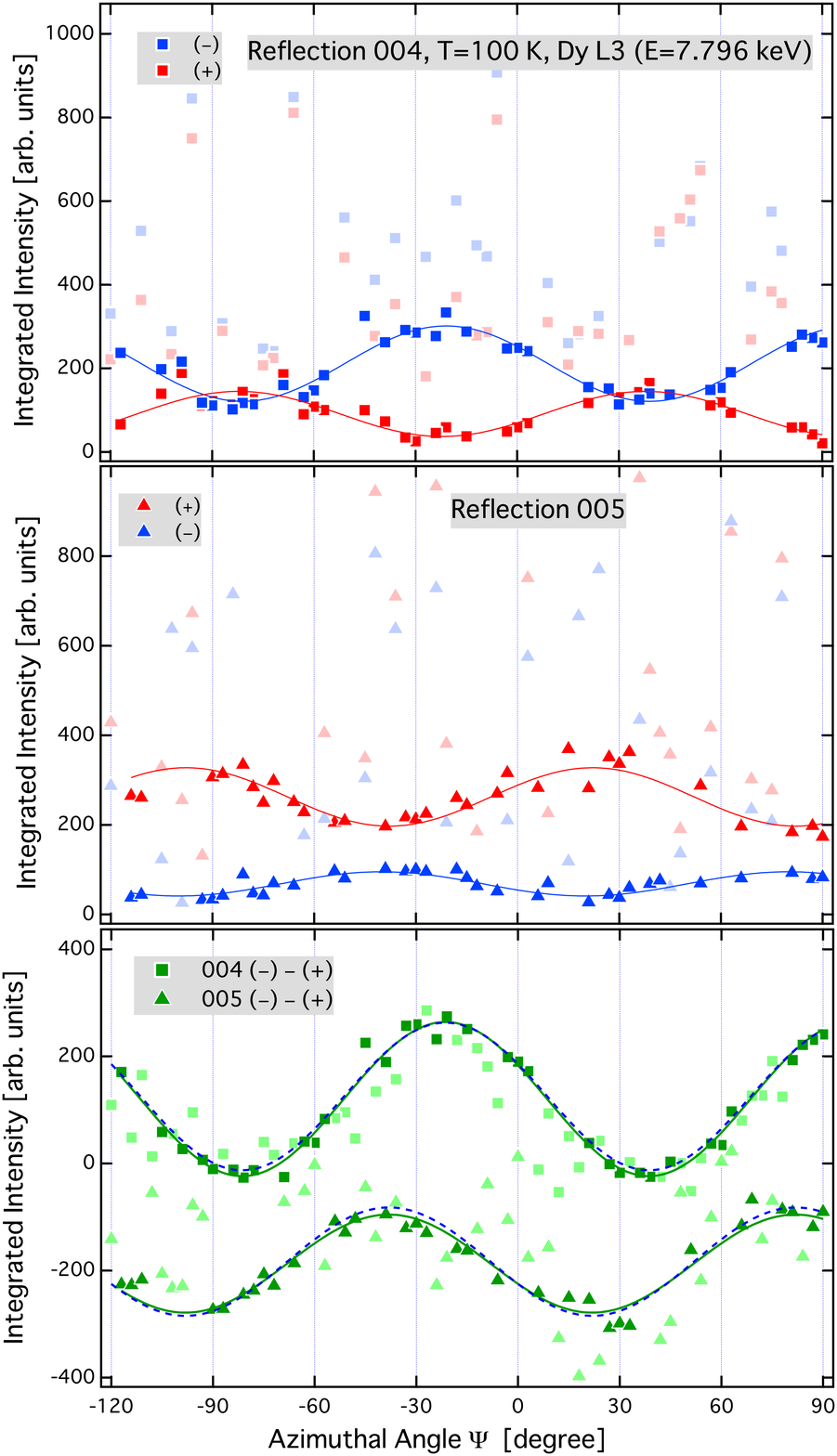}
\caption{
\label{fig:Dy-azimuth}
(Color online)  The integrated intensity of forbidden reflections 004
 (top panel) and 005 (middle panel) observed by $\omega$ scan at the Dy
 $L_3$ absorption  edge ($E=7.796$ keV) and $T=100$ K as a function of
 $\Psi$\@.  
 Red squares (triangles) show the integrated intensity measured by the
 incident beam with the $(+)$ helicity for reflection 004 (005) and
 blue squares (triangles) show that measured by the incident beam with
 the $(-)$ helicity for reflection 004 (005). 
 The bottom panel shows the difference intensity, $(-)-(+)$, between the 
 two helicity states.  The cosine curves are results of fit to the data
 with Eq.~\ref{eq:14} for the top and middle panels, and fit to the data
 with Eq.~\ref{equ:diff} for the bottom panel.  The points colored
 faintly are strongly influenced by the multiple scattering effect, and
 are removed for the fitting.
 The dash lines in the bottom panel are given from the result of the
 least square method described in the text.
 }
\end{figure}

\begin{figure}[htp]
\includegraphics[width=80mm]{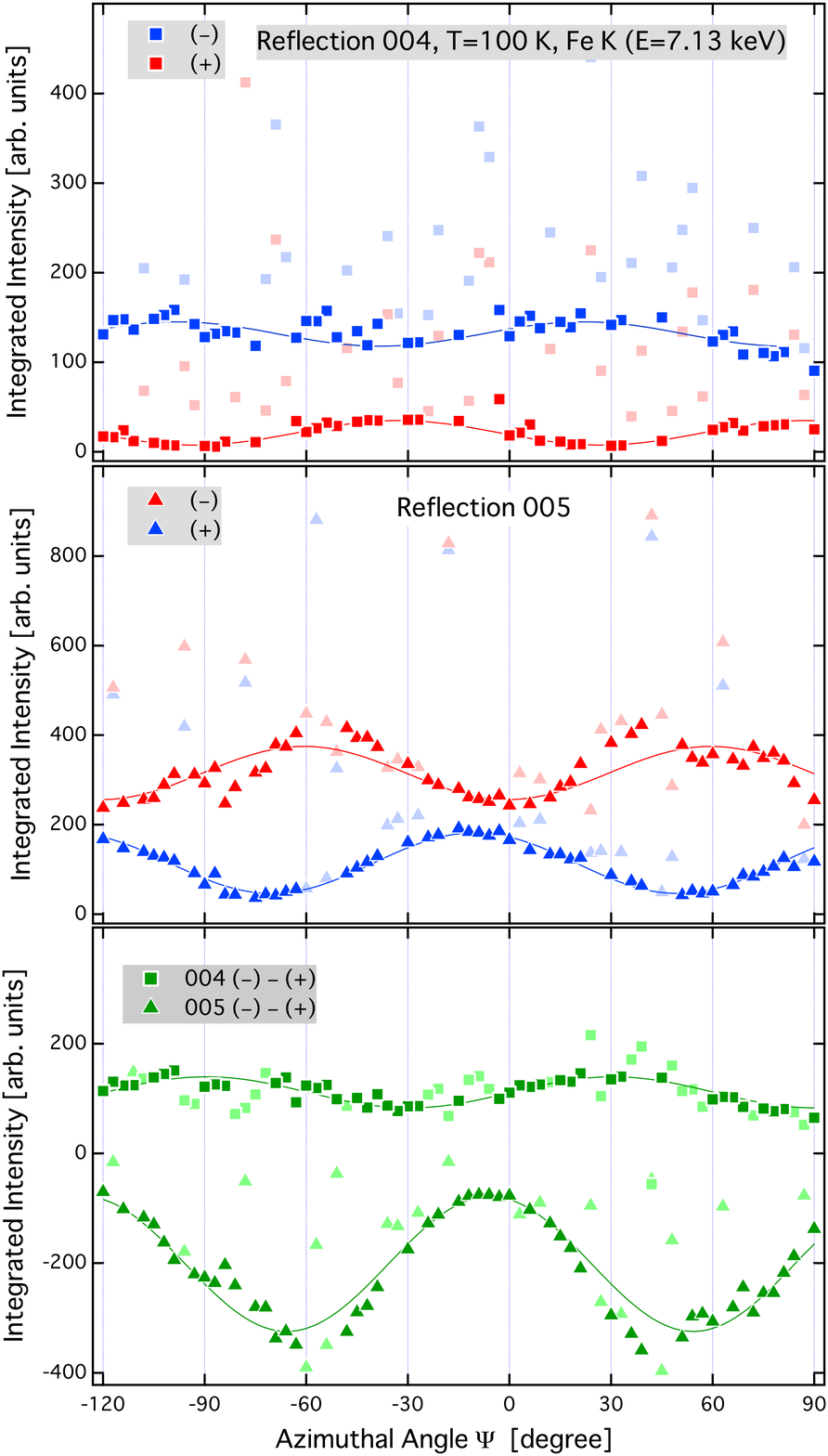}
\caption{
\label{fig:Fe-azimuth} 
 (Color online) The integrated intensity of forbidden reflections 004
 (top panel) and 005 (middle panel) observed by $\omega$ scan at the Fe
 $K$ absorption edge ($E=7.13$ keV) and temperature $T=100$ K as a
 function of $\Psi$\@. 
 Red squares (triangles) show the integrated intensity measured by the
 incident beam with the $(+)$ helicity for reflection 004 (005) and
 blue squares (triangles) show that measured by the incident beam with
 the $(-)$ helicity for reflection 004 (005). 
 The bottom panel shows the difference intensity, $(-)-(+)$, between the 
 two helicity states. 
The cosine curves are results of fit to the data with Eq.~\ref{eq:14}
 for the top and middle panels, and fit to the data with
 Eq.~\ref{equ:diff} for the bottom panel.  The points colored faintly
 are  strongly influenced by the multiple scattering effect, and are
 removed for the fitting. 
}
\end{figure}

Figures~\ref{fig:Dy-azimuth} and~\ref{fig:Fe-azimuth} show the
integrated intensity of reflections 004 and 005 for the $(+)$ and $(-)$ 
states and their difference intensity between two helicity 
states, $(-)-(+)$, at Dy $L_3$ ($E=7.796$ keV) absorption edge and at 
Fe $K$ ($E=7.13$ keV) absorption edge, respectively. All the data  
were measured at $T=100$ K\@.  The data are corrected with the
experimental Lorentz factor.  The absorption correction is not necessary
because the incoming and outgoing beam have the same angle to the sample
surface for both reflections.
The range of azimuth angle $\Psi$ for all scans is from $-120^\circ$ to
$90^\circ$.
 The bumpy structure of azimuth scan curves is due to the multiple
 scattering effect as described in Sec.~\ref{sec:absorption}. 
Taking the difference between the two helicity states eliminates most of 
the multiple scattering effect and gives rather smooth azimuth curves for
both reflections 004 and 005, and at both Dy $L_3$ and Fe $K$ absorption 
edges.  This is because the subtraction cancels out the multiple
scattering effect which is only very weakly dependent on the sign of
circular polarization. 
We find that these azimuth functions have a 120 degree periodicity
according to the trigonal crystal symmetry, that the intensity for the
$(-)$ helicity is higher than that for the $(+)$ helicity for reflection 
004 for both absorption edges, that this relation is reversed for
reflection 005.  This fact exactly indicates that the sample used in
this study belongs to the \emph{left-handed} space group $P3_221$, as we 
discuss later.  This handedness, of course, is in accord with that we
determined with soft x-ray diffraction at Dy $M_5$ absorption
edge.~\cite{Usui2014ooq}  
 
\subsection{Temperature dependence}
\label{sec:temperature}
\begin{figure}[htp]
\includegraphics[width=80mm]{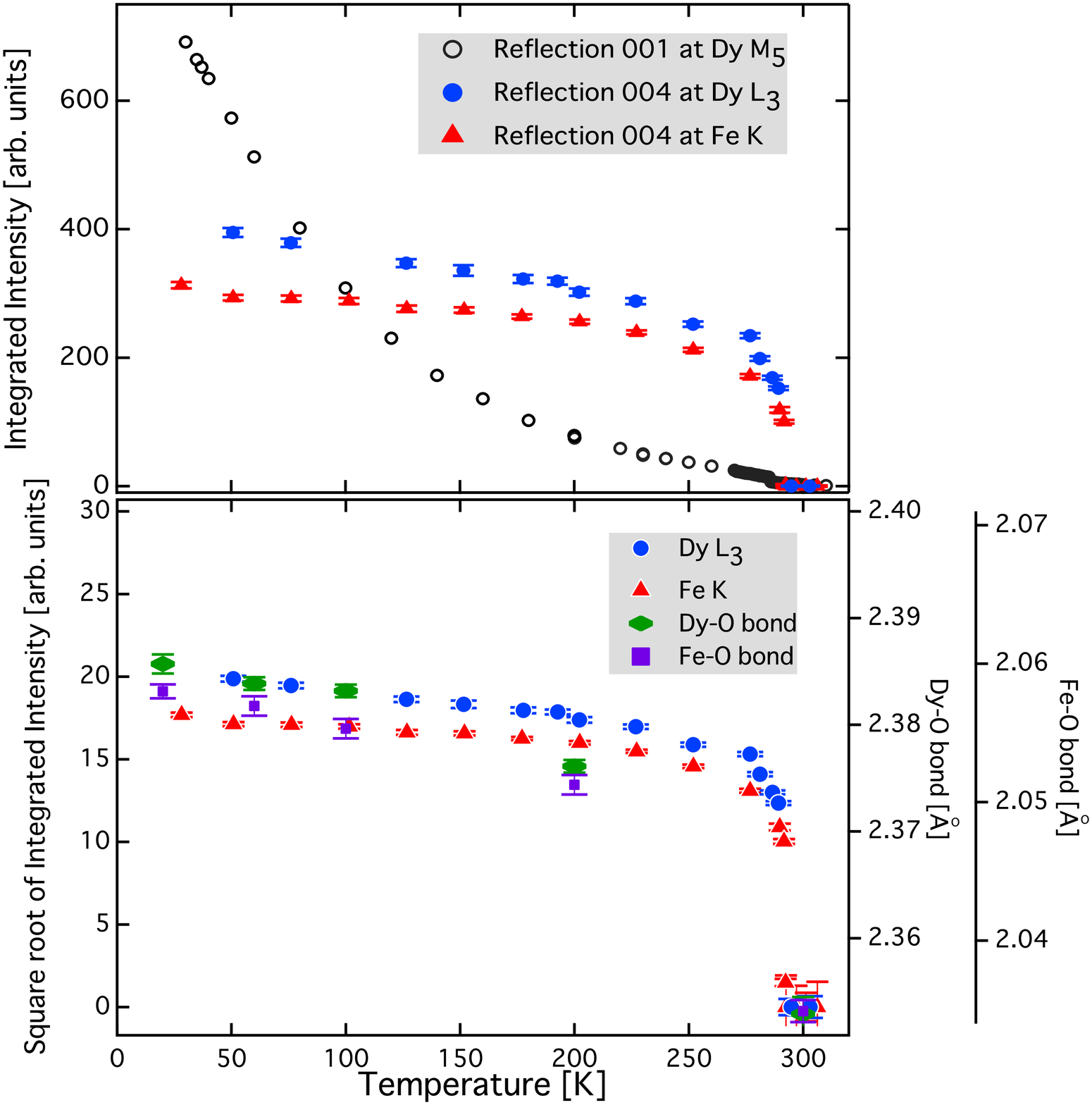}
\caption{
\label{fig:temperature}
(Color online) The temperature dependence of the integrated intensity of
 $\omega$  scans for reflection 004. Top: the integrated intensity of
 reflection  004 observed at Dy $L_3$ absorption edge (circles) and Fe
 $K$ absorption edge (triangles), and that for reflection 001 observed 
 at Dy $M_5$ absorption edge (open circles).~\cite{Usui2014ooq}
 The integrated intensity of reflection 001 is multiplied by a factor 
 to compare with the others. 
 Bottom: The square root of the integrated intensity of reflection 004 
 observed at Dy $L_3$ absorption edge (circles) and Fe $K$
 absorption edge (triangles) together with the bond length of Dy-O(4)
 (diamonds) and the bond length of Fe(2)-O(2$_1$) (squares).~\cite{Usui2014ooq}
}
\end{figure}

We observed the integrated intensity of forbidden
reflection 004 observed at Dy $L_3$ and Fe $K$ edges as a function of
temperature.  
Figure~\ref{fig:temperature} shows the integrated intensity of reflection
004 together with that of reflection 001 previously observed at Dy $M_5$
edge with a soft x-ray beam.~\cite{Usui2014ooq} 
We measured the intensity at several azimuth angles, where the multiple
scattering effect was expected to be negligible.
The data shown here is observed at one fixed azimuth angle. 
We find that the temperature evolution of the integrated intensities of
reflection 004 observed both at Dy $L_3$ and Fe $K$ edges is quite
different from that of reflection 001 observed at Dy $M_5$ edge. 
They show a rather monotonous increase on cooling after the jump just 
below the $T_S$ while that of reflection 001 observed at Dy $M_5$ edge
shows a steep increase towards lower temperatures.  
We discuss this feature comparing with the deformation of DyO$_6$
trigonal prism and FeO$_6$ octahedron in the next Section.

\section{Discussion and Analysis}\label{sec:discussion}
\subsection {The Dy $5d$ and Fe $4p$ quadrupole moments in the azimuth scans}\label{sec:quadrupole}
The intensity $I$ of the resonant diffraction is generally given with
the Stokes parameters,~\cite{Lovesey2005epo}
\begin{eqnarray}
I &= &\frac{1}{2} \left ( 1+P_3 \right )\left ( \left | 
G_{\sigma' \sigma} \right |^2 +  
\left | G_{\pi '\sigma} \right |^2 \right )  \nonumber \\ 
&+&\frac{1}{2} 
\left ( 1- P_3 \right )
\left (\left |  G_{\pi '\pi}  \right |^2 + \left | 
G_{\sigma '\pi} \right |^2 \right ) \nonumber \\
& + &  P_2 
\mathrm{Im}\left (  G_{\sigma '\pi}^*  G_{\sigma '\sigma} + 
G_{\pi' \pi}^*   
G_{\pi '\sigma} \right ) \nonumber \\
& + &  P_1 
\mathrm{Re}\left (  G_{\sigma '\pi}^*  G_{\sigma '\sigma} +  
G_{\pi '\pi}^*  G_{\pi '\sigma} \right ). 
\label{equ:cross-section} 
\end{eqnarray}
Here $G_{\alpha '\beta}$ is the total resonant scattering amplitude, and
$\alpha'$ and $\beta$ are the polarization state of the diffracted and 
incident x-ray beam, respectively.
The third term including $P_2$ represents the interference between
$\sigma$ and $\pi$ components in the resonant scattering process for the
circularly polarized x rays.
It plays a crucial role for determination of the chirality by changing
the sign coupling with the helicity of the x-ray beam.  

We use the atomic multipoles, $\left \langle T^{K}_Q \right \rangle$, 
the expectation value of the spherical tensor, to express the scattering
amplitude $G_{\alpha '\beta}$ as described in
Ref.~\onlinecite{Lovesey2005epo}. 
This method is so powerful that we discuss the components of the
multipole moment ordered in materials. 
The total scattering amplitude is composed of  possible resonant events,
\begin{eqnarray}
\label{equ:G}
 G_{\alpha '\beta}   &=& \sum_k r^{(k)}\frac{F^{(k)}_{\alpha '\beta}}
 {\hbar \omega-\Delta +\frac{i}{2} \Gamma}\\ \nonumber
               &=& \sum_k d^{(k)}(E)F^{(k)}_{\alpha '\beta} \; ,  
\end{eqnarray}
where $k$ represents individual resonant events like $E1E1$, $E1E2$ etc.,
$\Delta$, $\Gamma$, and  $r^{(k)}$ represent the resonant energy,
its width, and the mixing parameter, respectively, and the scattering 
length $F_{\alpha' \beta}$ is 
\begin{eqnarray}
 F_{\alpha '\beta}   = \sum_K X^{(K)}_{\alpha '\beta}D^{(K)}\Psi^{(K)}. 
\label{equ:F} 
\end{eqnarray}
Here $\Psi^{(K)}$ represents the unit-cell-structure-factor tensor of
rank $K$ which is the sum of atomic multipoles related to the resonant
process, and $X^{(K)}_{\alpha '\beta}$ describes the conditions of the
incident and the diffracted beam.  Orientation of the crystal, with
respect to states of polarization and the plane of scattering, is
accomplished by a rotation matrix $D^{(K)}$. 
Note that the atomic scattering length expressed by
Eq.~\ref{equ:resonance} is summed up in the unit-cell structure factor 
$\Psi^{(K)}$  including the geometric factor $X^{(K)}_{\alpha
'\beta}D^{(K)}$, and that the tensorial character is emerged with the
x-ray polarization.  

As far as the Dy $4f$ and $5d$ quadrupole moments are concerned, the
structure factor for the enantiomorphic space-group pair $P3_121$
(\#152) and $P3_221$ (\#154) described
in Ref.~\onlinecite{Lovesey2008rdc} is applicable to \DFBO\ because the 
Dy site is at the special position $3a$
(multiplicity 3 and Wyckoff letter $a$), which is the same as the Si
site in low-quartz.
For the space groups \#152 and \#154, Dy ions locate at
($x$, 0, $\pm\frac{1}{3}$), (0, $x$, $\mp\frac{1}{3}$),
and ($-{x}$, $-{x}$, 0) in a unit cell.  
Here the upper and lower sign represent the space groups \#152 and
\#154, respectively.
The unit-cell structure factor is the sum of the atomic multipoles,
each of which is defined at each Dy ion.
First, we define the atomic multipole
 $\left \langle T^{K}_Q \right \rangle$ at the ion ($-{x}$, $-{x}$, 0),
 which is common for both space groups,  with a right-handed Cartesian
 axes $(\xi\eta\zeta)$.  
Here the $\zeta$ axis is parallel to the $c$-axis, and the $\xi$ axis
parallel to the [$1$, $1$, 0] axis, which encloses an angle of 30
degrees with the reciprocal $a^*$ axis. 
The atomic multipoles of the other Dy ions at
($x$, 0, $\pm\frac{1}{3}$), and (0, $x$, $\mp\frac{1}{3}$) are given
 as $\left \langle T^{K}_Q \right \rangle \exp(2\pi i Q/3)$, and 
$\left \langle T^{K}_Q \right \rangle \exp(-2\pi i Q/3)$, respectively,
with a rotation along the $\zeta$ axis by $+120$, $-120$ degrees,
respectively. 
The unit-cell structure factor for reflection $00l$ is 
\begin{equation}
\begin{aligned}
\Psi_Q^{K} =  \left \langle T^{K}_Q \right \rangle 
  & \left \{ 1 +\mathrm{e}^{2\pi i Q/3} \mathrm{e}^{2\pi i  (\pm l/3) } \right .  \\
  & +\left . \mathrm{e}^{-2\pi i Q/3} \mathrm{e}^{2\pi i  (\mp l/3) }  \right \}.
\end{aligned}
\end{equation}
Here we find the selection rule for reflection $00l$ that $l+Q=3n$
for space group $P3_121$ and $l-Q=3n$ for space group $P3_221$, 
where $n$ is an integer.
The handedness emerges as the sign of $Q$\@.

The most probable resonant process at the Dy at $L_3$ absorption
edge is the $E1E1$ event, the resonance between $2p_{\frac{3}{2}}$ and
$5d$.
As we discuss later, the asymmetric process $E1E2$ is negligible.
The atomic multipole related to the $E1E1$ event is
$\left \langle T^{K}_Q \right \rangle$ with rank $K=2$,
in the other words, the quadrupole.
There are five independent components for the quadrupole moment with relations,
$\left \langle T^{2}_{+1}\right \rangle' =-\left \langle T^{2}_{-1}\right \rangle'$,
$\left \langle T^{2}_{+1}\right \rangle'' =\left \langle T^{2}_{-1}\right \rangle''$,
$\left \langle T^{2}_{+2}\right \rangle' =\left \langle T^{2}_{-2}\right \rangle'$, and
$\left \langle T^{2}_{+2}\right \rangle'' =-\left \langle T^{2}_{-2}\right \rangle''$.
Note that the atomic multipole $\left \langle T^{K}_Q \right \rangle$  is 
a complex number defined as
$\left \langle T^{K}_Q \right \rangle =\left \langle T^{K}_Q \right 
\rangle ' + i \left \langle T^{K}_Q \right \rangle'' $
with 
$\left \langle T^{K}_Q \right \rangle^*=(-1)^Q\left \langle T^{K}_{-Q} \right \rangle$.
The five independent components are real numbers; $\left \langle T^{2}_0 \right \rangle'$, 
$\left \langle T^{2}_{+1}\right \rangle'$, $\left \langle T^{2}_{+1}\right \rangle''$,
$\left \langle T^{2}_{+2}\right \rangle'$, and $\left \langle T^{2}_{+2}\right \rangle''$, 
each of which corresponds 
to $3\zeta^2-r^2$, $\zeta\xi$, $\eta\zeta$, $\xi^2-\eta^2$, and
$\xi\eta$ components, respectively. 
Among the five components, 
$\left \langle T^{2}_{+1} \right \rangle'$,
and $\left \langle T^{2}_{+2} \right \rangle''$
are zero for the Dy ions because the site $a$ is on the two fold axis 
which gives the relation
 $\left \langle T^{K}_Q \right \rangle = (-1)^K\left \langle T^{K}_{-Q} \right \rangle$.
The selection rule, $l \pm Q =3n$, for space-group forbidden reflections $00l$ ($l \neq 3n$)
excludes the $\left \langle T^{2}_0 \right \rangle'$ component.
Therefore, as far as $E1E1$ process is concerned at Dy $L_3$ edge,
the intensity of space-group forbidden reflections is described by only 
two components $\left \langle T^{2}_{+2} \right \rangle'$,  
and $\left \langle T^{2}_{+1} \right \rangle''$. 
Note that the mirror operation between
enantiomorphic space-group pair \#152 and \#154 gives
\begin{eqnarray}
\left \langle T^{2}_{+1} \right \rangle''_{\#152} &=&- \left \langle T^{2}_{+1} \right \rangle''_{\#154} \label{eq:8} \\
\left \langle T^{2}_{+2} \right \rangle'_{\#152} &=& \left \langle T^{2}_{+2} \right \rangle'_{\#154} \label{eq:9}\: .
\end{eqnarray}
Eq.~\ref{eq:9} shows that the component on the basal plane is common for
both of the enantiomorphic space-group pair, 
while Eq.~\ref{eq:8} shows that the component on the $\eta\zeta$ plane
has the opposite sign to each other.
The four amplitudes~\cite{Lovesey2008rdc} for space-group forbidden
reflections $00l$, $l=3n+\mu$ ($\mu=\pm 1$) for the parity-even event
$E1E1$ at Dy $L_3$ absorption edge are 
\begin{eqnarray}
G_{\sigma'\sigma}&=& \frac{3}{2} \left <T_{+2}^2 \right >' \mathrm{e}^{i\phi_a} , \\
G_{\pi'\pi}&=&\frac{3}{2} \left <T_{+2}^2 \right >' \mathrm{e}^{i\phi_a} \sin ^2 \theta , \\
G_{\pi'\sigma}&=& \frac{3}{2} i \lambda  
\left \{    \left <T_{+2}^2 \right >' \mathrm{e}^{i\phi_a} \sin \theta \right . \nonumber \\
 &  & +\left . \left <T_{+1}^2\right >'' \mathrm{e}^{i\phi_b} \cos \theta \mathrm{e}^{3 i \lambda \Psi}  \right \}, \\ 
G_{\sigma'\pi}&=& \frac{3}{2} i \lambda  
\left \{ - \left <T_{+2}^2 \right >' \mathrm{e}^{i\phi_a} \sin \theta  \right . \nonumber \\
 & &+  \left . \left <T_{+1}^2\right >'' \mathrm{e}^{i\phi_b} 
 \cos \theta \mathrm{e}^{3 i \lambda \Psi}  \right \}.
 \end{eqnarray}
Here Bragg angle $\theta$ and the azimuth angle $\Psi$ are shown
and defined in Fig.~\ref{fig:geometry}.
The phase factors $\mathrm{e}^{i\phi_a}$ and ${e}^{i\phi_b}$ emerge 
from Eq.~\ref{equ:G}, and depend on the x-ray energy differently to 
each other.~\cite{Joly2014cbp,Takahashi2015}
Here we introduce two parameters $\nu=\pm1$ and $\lambda=\pm1$.
The parameter $\nu$ denotes the crystal handedness, namely
$\nu=+1$ for the right-handed space group \#152, and $\nu=-1$
for the left-handed space group \#154, and the parameter $\lambda$
denotes the product of $\mu$ and $\nu$, $\lambda=\mu \nu$.
Note that the rotating direction of $\Psi$ is opposite to that defined
in Ref.~\onlinecite{Lovesey2008rdc}.

The intensity $I$ for the $E1E1$ event at Dy $L_3$ absorption edge is,  
\begin{eqnarray}
I &=& I_0 + I_1 \cos  3 \left( \lambda \Psi +  \phi \right) , \label{eq:14} \\
I_0 &=& \frac{1}{2} \left \{  \left( 1+ \sin^2\theta \right)^2 T_a^2 
+ 2  \cos^2 \theta \,T_b^2  \right . \nonumber \\
 & & \left . + P_3 \left ( 1+\sin^2 \theta \right ) \cos^2 \theta \, T_a^2  \right \}  
 \nonumber \label{eq:15} \\
& & +  \lambda P_2  \sin \theta \left ( 1+ \sin^2 \theta \right) T_a^2 , \\
I_1 &=&  \left  (2P_3  \sin \theta \cos \theta - \lambda P_2 \cos ^3 \theta \right ) 
T_a T_b. \label{eq:16}
\end{eqnarray}
Here we define the parameters $\phi=\frac{1}{3}\left ( \phi_b-\phi_a \right )$, 
$T_a=\frac{3}{2} \left <T_{+2}^2 \right >'$, 
$T_b=\frac{3}{2}\left <T_{+1}^2\right >''$, 
and we presume $P_1=0$ for simplicity.
Eqs.~\ref{eq:14}, \ref{eq:15}, and~\ref{eq:16} are the same as those 
of Ref.~\onlinecite{Lovesey2008rdc} in the $E1E1$ transition except 
the presence of the shift $\phi$ in $\Psi$\@.
The intensity is a three-fold periodic function of the azimuthal angle 
$\Psi$.
The circular polarization, $P_2$, is multiplied by $\lambda$, indicating 
that reversing the crystal chirality $\nu$ together with reversing the 
sign of circular polarization does not change the intensity.
Only the shift $\phi$ in $\Psi$ changes its sign with $\lambda$ 
according to Eq.~\ref{eq:14}.
The relation of Eq.~\ref{eq:8} is absorbed in $T_b$ itself in 
Eq.~\ref{eq:16}.
By extracting it, we find 
\begin{eqnarray}
I_1 =  \left  (2\nu P_3  \sin \theta \cos \theta -  \mu P_2 \cos ^3 \theta \right ) T_a T_b, \label{eq:17}
\end{eqnarray}
with $T_b=T_{b(\#152)}=-T_{b(\#154)}$.
Eq.~\ref{eq:17} gives relations 
(i) $I_1(+\nu, \pm P_2) = -I_1(-\nu, \mp P_2)$ for the same reflection 
index, (ii) when $P_3=0$, $I_1(+\nu, \pm P_2) = I_1(-\nu, \pm P_2)$ for 
the same reflection index, and (iii) when $P_3=0$, $I_1$ for reflection 
indices $\mu=+1$ and $\mu=-1$ has the opposite sign to each other for 
the same values of $\nu$ and $P_2$.
The previous observations 
for \DFBO\ and low-quartz,~\cite{Usui2014ooq,Tanaka2008rho,Joly2014cbp} 
support the relation (i), and also suggest the relation (ii) for the 
small value of $P_3$.
The relation (iii) has been confirmed in the data sets of Te and 
AlPO$_4$.~\cite{Tanaka2010doa,Tanaka2010dsc}
The difference intensity between two helicity states ($P_2=-1$ and 
$P_2=+1$) is 
\begin{equation}
          I(-)-I(+) = A + B \cos 3( \lambda  \Psi +\phi),
\label{equ:diff}
\end{equation}
with $A=-2 \lambda \sin \theta \left ( 1+ \sin^2 \theta \right)T_a^2$, 
and $B=2 \mu \cos ^3 \theta \, T_a T_b$.
Note that there are only two quadrupole components, $T_a$ and $T_b$, for 
the intensity at Dy $L_3$ absorption edge.

Analyzing the Fe $4p$ quadrupole moments observed at the Fe $K$ absorption
edge is more complicated because there are nine ions in a unit cell:
three of them, Fe(1) are at the special position $3a$ (multiplicity 3,
Wyckoff letter $a$) and the other six, Fe(2) are at the general position
$6c$ (multiplicity 6, Wyckoff letter $c$) in the low-$T$ phase.  
For $00l$ reflection, six independent $4p$ quadrupole components contribute in total:
they are 
two  quadrupole components of three Fe(1) ions,
\begin{eqnarray}
Q_{\xi^2-\eta^2}^{(1)}=\left \langle  T_{+2}^2 \right \rangle '^{(1)} e^{i p_1}, \nonumber \\ 
Q_{\eta\zeta}^{(1)}=\left \langle  T_{+1}^2 \right \rangle ''^{(1)} e^{i q_1}, \nonumber
\end{eqnarray}
and  
four quadrupole components of six Fe(2) ions,
\begin{eqnarray}
Q_{\xi^2-\eta^2}^{(2)}=\left \langle  T_{+2}^2 \right \rangle '^{(2)} e^{i p_2}, \nonumber\\
Q_{\xi\eta}^{(2)}        =\left \langle  T_{+2}^2 \right \rangle ''^{(2)} e^{i q_2}, \nonumber \\
Q_{\eta\zeta}^{(2)}    =\left \langle  T_{+1}^2 \right \rangle ''^{(2)} e^{i r_2}, \nonumber \\ 
Q_{\zeta\xi}^{(2)}      =\left \langle  T_{+1}^2 \right \rangle '^{(2)} e^{i s_2}. \nonumber
\end{eqnarray}
The components are defined with the local coordinate ($\xi\eta\zeta$) in the same way as Dy components. 
Each component is a complex number and has an independent phase factor according to Eq.~\ref{equ:G}\@.  
When $P_1=0$, we obtain the intensity of forbidden reflection $00l$ at Fe $K$ edge,
 \begin{eqnarray}
 I &=& I_0  + I_1 \cos 3 \left( \lambda  \Psi + \phi \right) 
    \label{eq:FeI} \;, \\  
 I_0 &=& \frac{1}{2} \left \{ \left(1+\sin^2\theta\right)^2 T_{\alpha}^2 + 2 \cos^2\theta \: 
 T_{\beta}^2 \right . \nonumber \\  
       & & +  \left . P_3  \left(1+\sin ^2 \theta\right) \cos ^2 \theta \: T_{\alpha}^2 \right \} \nonumber \\
       & &       +\lambda P_2 \sin \theta \left(1+\sin^2\theta\right) T_{\alpha}^2   \;,  \label{eq:FeI0}\\
 I_1 &=& \left ( 2\nu P_3  \sin \theta \cos \theta - \mu   P_2 \cos^3\theta \right ) T_{\alpha} T_{\beta} \;,  \label{eq:FeI12} 
  \end{eqnarray}
with
\begin{subequations}
\label{FaFb}
\begin{eqnarray}
T_{\alpha} e^{i\phi_{\alpha}} &=& 3 \left \{  \frac{(-1)^l}{2} Q_{\xi^2-\eta^2}^{(1)}+ \cos \rho \, Q_{\xi^2-\eta^2}^{(2)}   \right . \\
& & \left . - \mu  \sin \rho \, Q_{\xi\eta}^{(2)}  \right \}, \nonumber  \\
T_{\beta} e^{i\phi_{\beta}}  &=& 3 \left \{  \frac{(-1)^l}{2}  Q_{\eta\zeta}^{(1)}  
+  \cos \rho \, Q_{\eta\zeta}^{(2)} \right . \nonumber \\  
& & \left . + \mu  \sin \rho \, Q_{\zeta\xi}^{(2)}   \right \}.
\end{eqnarray}
\end{subequations}
Here $\phi = \frac{1}{3}(\phi_{\beta} - \phi_{\alpha})$
and $ \rho = 2 \pi l \left(\frac{1}{6} -\delta  \right) $.
We define $\delta$ as $0.32333 = 1/3 - \delta$, and $\delta = 0.010003$
for one of the Fe(2) ions, which is located at (0.54906, 0.335907, 0.32333) in a unit cell.
Note that the parameters $T_{\alpha}$ and $T_{\beta}$ are real numbers. 
As we see, Eq.~\ref{eq:FeI0} and Eq.~\ref{eq:FeI12} have the same form
as Eq.~\ref{eq:15} and Eq.~\ref{eq:17} do, respectively, by 
replacing $T_a$ and $T_b$ with $T_{\alpha}$ and $T_{\beta}$.
Therefore we employ the same functions to analyze the azimuth-angle-scan
data sets for Dy $L_3$ and Fe $K$ absorption edges.

We fit the data sets of the azimuth angle scans shown in 
Figs.~\ref{fig:Dy-azimuth} and~\ref{fig:Fe-azimuth} 
for the integrated intensity of two helicity states ($P_2<0$ and 
$P_2>0$) with Eq.~\ref{eq:14}, and for difference intensity, $I(-)-I(+)$ 
with Eq.~\ref{equ:diff}, together with the origin shift $\phi$ in $\Psi$.
In practice, the fitting was carried out with selected data points to 
make a cosine envelop curve for each data set, otherwise, the 
multiple scattering prevents us from fitting properly.
Some arbitrariness is inevitable for this selection, however, each 
curve shown in Figs.~\ref{fig:Dy-azimuth} and~\ref{fig:Fe-azimuth} traces 
an envelop shape in each data set very well.
The results of fit are summarized in Table.~\ref{tab:azimuth}. 
The points colored faintly in both figures are removed for the fitting.  
\begin{center}
\begin{longtable*}[c]{c|c|ccc|ccc}
\caption{
The values of the parameters as results of fit to the data sets
of the azimuth angle scans for reflections 004 and 005 shown in
Figs.~\ref{fig:Dy-azimuth} and~\ref{fig:Fe-azimuth}.
Two helicity states ($P_2<0$ and $P_2>0$) are represented by $(-)$ and
 $(+)$, respectively.
 The data sets are fit by Eqs.~\ref{eq:14} with parameters,  $I_0$,
 $I_1$, and $\phi$, and the subtracted data, $I(-)-I(+)$, are fit by
 Eq.~\ref{equ:diff} with parameters, $A$, $B$, and $\phi$.
}
\label{tab:azimuth} \\
\multicolumn{2}{c|}{} & \multicolumn{3}{c|}{Dy $L_3$ edge} & \multicolumn{3}{c}{Fe $K$ edge} \\
\hline
 &  Helicity ($P_2$)  & $I_0$($A$) & $I_1$($B$) & $\phi$ & $I_0$($A$) & $I_1$($B$) & $\phi$\\
\hline
\hline \endfirsthead
\multirow{3}{*}{004}
& $(-)$ & $211.2\pm4.2$ & $-90.1\pm5.6$ & $39.1\pm1.3$  &  $131.3\pm1.4$ & $-14.7\pm1.9$ & $-33.7\pm2.8$\\
\cline{2-8}
& $(+)$ & $90.9\pm3.7$ & $53.9\pm5.0$ & $37.8\pm1.9$ & $21.2\pm0.5$  & $13.7\pm0.8$ & $-32.9\pm1.2$\\
\cline{2-8}
& $(-)-(+)$ & $120.3\pm3.3$ & $-143.9\pm3.8$ & $38.6\pm0.6$ & $109.1\pm1.5$ & $-26.5\pm1.9$ & $-33.7\pm1.7$\\
\hline \hline
\multirow{3}{*}{005}& $(-)$  & $68.2\pm2.3$ & $27.1\pm3.3$ & $40.1\pm2.4$ & $114.0\pm1.5$ & $67.7\pm2.2$ & $10.2\pm0.6$ \\
\cline{2-8}
& $(+)$ & $262.3\pm5.2$ & $-65.2\pm6.8$ & $37.9\pm2.2$ & $315.4\pm4.9$ & $-59.4\pm6.9$ & $0.6\pm2.2$\\
\cline{2-8}
&  $(-)-(+)$  & $-187.1\pm3.8$ & $91.6\pm5.0$ & $38.1\pm1.2$ & $-201.4\pm3.2$ & $123.0\pm5.8$ & $5.7\pm0.9$\\
\hline \endlastfoot
\end{longtable*}
\end{center}

We find that parameter $A$ is positive for reflection 004 and is
negative for reflection 005 at Dy $L_3$ absorption edge, which indicates
that $\nu=-1$, ie the sample we observed belongs to the left-handed
space group \#154. 
This is consistent with the previous observation at the Dy $M_5$
edge.~\cite{Usui2014ooq}
Likewise, the parameter $A$ for the data observed at the Fe $K$  edge is
positive and the same discussion is applicable to the handedness of the
Fe structure in this sample.
Of course, the handedness of the Dy structure and the Fe structure is
coincident in the same space group.
Moreover, we find that the relation (iii) for $I_1$ in Eq.~\ref{eq:17}
is confirmed between reflections 004 and 005 at both absorption edges,
and that the origin shift $\phi$ in $\Psi$ is almost unchanged for both 
two helicity states $(-)$, and $(+)$ at the Dy $M_5$ edge, which 
evidences that the $E1E1$ process is good enough to demonstrate the
experimental data at the Dy $L_3$ edge. 
The discrepancy in $\phi$ about 12 degrees between two helicity states
has been found in the study of low-quartz.~\cite{Tanaka2008rho}  This
discrepancy has been discussed in terms of the birefringence phenomenon,
higher-order transition processes like $E1E2$, or the x-ray polarization 
itself.~\cite{Joly2014cbp}  Finally, it has been concluded 
that a tiny but non-negligible off-diagonal polarization $P_1$ possibly
causes such discrepancy, and that the effects of higher-order transition
processes and  the birefringence should be very small, judging from the
\emph{ab initio} simulation and the XANES experimental data.  
Present data sets do not show any obvious discrepancy in $\phi$ between
two helicity states for both reflections 004 and 005 at Dy $L_3$ edge.
The same discussion is applicable to the origin shift $\phi$ in $\Psi$
for the data observed at the Fe $K$ edge.
The values of $\phi$ is almost the same
for both  two helicity states $(-)$, and $(+)$ for reflection 004,
although there is some deviation for reflection 005.
Note that in case of Fe $K$  edge, 
the origin shifts $\phi$ for reflection 004 and 005 is not necessarily
coincident.
Eq.~\ref{FaFb} shows that
six independent Fe $4p$ quadrupole components are summed up in two parameters,
$T_{\alpha}$ and $T_{\beta}$, both of which depend on the reflection index with the
interference between these sites. 
Consequently, the scattering amplitude does not necessarily have the
same value of $\phi$ for reflections 004 and 005, although the circular
polarization does not change $\phi$.  

Let us discuss about the quadrupole moments with the experimental
data.  Analyzing the Dy $5d$ quadrupole moments is easy because there
are only two parameters,
$\left <T_{+2}^2 \right >'$ and $\left <T_{+1}^2\right >''$ in four
equations.  
On the other hand, it is not easy for the Fe $4p$ quadrupole moments,
because there are six parameters summed up in Eq.~\ref{FaFb}.
We escape to deduce anything about the Fe $4p$ quadrupoles because
of the complexity.

Hereafter we discuss the Dy $5d$ quadrupoles. 
In case of pure circularly polarized states ($P_2=-1$ and $P_2=+1$),
Eq.~\ref{eq:14} is applicable to solve the parameters,  however, in
practice, the experimental Stokes parameters described in
Table~\ref{tab:polarization} are not simple because $P_3$ has
non-negligible values. 
In order to solve the parameters, we applied the least square method 
described in Appendix~\ref{appendix:solve} for the equations emerged
from reflections 004 and 005.
The results are
$\left | \left <T_{+2}^2 \right >' \right | =7.77\pm 0.06$ and
$\left | \left <T_{+1}^2\right >'' \right |=5.20\pm 0.15$ for the Dy
$5d$ quadrupole moment. 
The ratio is
$r=\left | {\left<T_{+1}^2\right >''} / {\left <T_{+2}^2 \right >' } \right |=0.67\pm 0.02$.
The dash lines at the bottom panel of Fig.~\ref{fig:Dy-azimuth}
reproduced with these values are quite similar to the lines obtained
from the independent fits by Eq.~\ref{equ:diff}. 
Note that the absolute values are meaningless, only the ratio has a
meaning, and that we cannot determine the signs of these components
because the phase shift $\phi$ has arbitrariness of 60 degrees (a half
period in $\Psi$), accompanied with the sign of $I_1$.
The ratio $r=0.67\pm 0.02$ for the Dy $5d$ quadrupole moment is smaller
than the ratio, $r=1.20$, for the Dy $4f$ moment observed at $T=100$
K,~\cite{Usui2014ooq} in other words, the ($\eta\zeta$) component is
smaller than that of the ($\xi^2-\eta^2$) in Dy $5d$  moment comparing
with the Dy $4f$ moment.
In the following subsection, we expect that the Dy $5d$ quadrupole is
strongly coupled with the deformation of DyO$_6$ trigonal prism as shown
in Fig.~\ref{fig:deformation}. 

\subsection {The deformation of DyO$_6$ trigonal prism and FeO$_6$ octahedron as a function of temperature}\label{sec:deformation}

According to the previous crystal structure analysis,~\cite{Usui2014ooq} 
the structural phase transition lowers the site symmetry of Dy, which is
on the three-fold rotation axis in the high-$T$ $R32$ phase, 
with a translational shift along the two-fold axis and a deformation of
the DyO$_6$ trigonal prism: the Dy-O(4) bond elongates and the Dy-O(3)
and Dy-O(7) bonds shrink at $T<T_S$ while the six oxygen atoms are
equidistant from the central Dy ion at $T>T_S$.  Accordingly, the site
symmetry of Dy changes from $D_3$ to $C_2$, losing the three-fold
symmetry in the low-$T$ $P3_121$ or $P3_221$ phase.
Here the number after the chemical symbol indicates the site position in 
the atomic coordinate determined by the single-crystal x-ray diffraction  
measurements, of which information is provided in the supplementary
information of Ref.~\onlinecite{Usui2014ooq}.   

\begin{figure}[htp]
\includegraphics[width=80mm]{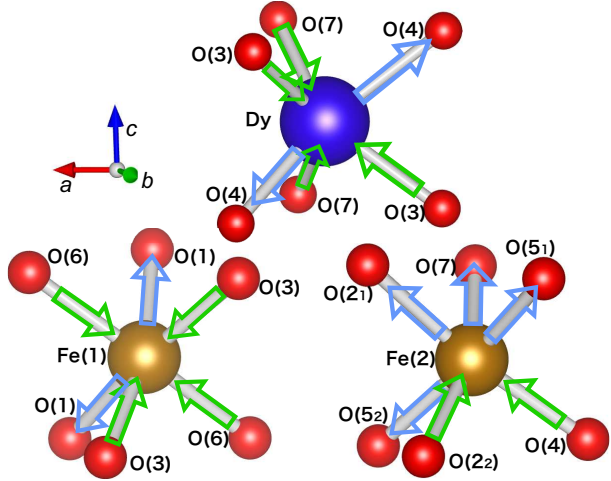}
\caption{
\label{fig:deformation}
 (Color online) An image of deformation in a DyO$_6$ trigonal prism and two FeO$_6$
 octahedra in the low-$T$ $P3_221$ phase due to the structural phase 
 transition at $T_S$. This image is drawn by VESTA.~\cite{Momma2011vft}
 Here the Fe(1) ion locates at $3a$ site and the Fe(2) ion locates at $6b$
 site in the low-$T$ $P3_221$ phase.  The arrows  show the direction of
 the elongation or the shrinkage. Pair bonds Fe(2)-O(2) lose the symmetry and 
 that  the Fe(2)-O($2_1$) bond elongates and the other Fe(2)-O($2_2$)
 bond shrinks.    
}
\end{figure}

Likewise, the FeO$_6$ octahedron deforms in the low-$T$ phase.  
The Fe ions are located at site $9d$, which is at the site symmetry $C_2$: one of the
three two-fold axes perpendicular to the $c$ axis, in the high-$T$ $R32$
phase.
As described in the previous section,
in the low-$T$ phase, the nine Fe ions are separated into two groups: 
three Fe(1) ions are at the special position $3a$, remaining on one of three two-fold axes, 
and six Fe(2) ions are at the general position $6c$.  
Fig.~\ref{fig:deformation} illustrates the deformation of the DyO$_6$ 
trigonal prism and Fe(1)O$_6$, Fe(2)O$_6$ octahedrons.
Here pair bonds Dy-O(3), Dy-O(4), and Dy-O(7) in the DyO$_6$, and
Fe(1)-O(1), Fe(1)-O(3), and Fe(1)-O(6) in Fe(1)O$_6$ are equidistant,
respectively, in the low-$T$ phase so as that Dy and Fe(1) ions are
kept to be on the two-fold axes perpendicular to the $c$ axis, whereas
all six bonds Fe(2)-O in Fe(2)O$_6$ octahedron are different to each
other. 

We show the deformation of the DyO$_6$ trigonal prism and Fe(1)O$_6$,
Fe(2)O$_6$ octahedrons as a function of temperature by taking two
typical bonds, Dy-O(4) and Fe(2)-O($2_1$) in the bottom panel of
Fig.~\ref{fig:temperature}  
together with the square root of the integrated intensity of reflection 
004 observed at two absorption edges, Dy $L_3$ and Fe $K$\@.
As seen, we find that the square root of the integrated intensity shows 
a monotonous change with temperature as well as the lengths of two
bonds,  Dy-O(4) and Fe(2)-O($2_1$), indicating that the  Dy $5d$
quadrupole moment strongly couples with ligand oxygen atoms. 
On the other hand,  the integrated intensity of reflection 001 at Dy
$M_5$ edge shows a drastic increase with decreasing temperature,
indicating that the development of the Dy $4f$ quadrupole moment  
is independent of the deformation of the DyO$_6$ trigonal prism.  
This development is possibly caused by the population at the respective
sub-levels in eight Kramers doublets in the $^6H_{\frac{15}{2}}$ state
of Dy$^{+3}$ $4f$ electrons.
Note that the splitting energy between the first and the second
sub-levels is estimated to be  
$15 \sim 20  $ cm$^{-1}$.~\cite{Popova2008m,Stanislavchuk2007iib,Volkov2008mpd,Malakhovskii2013moa}

\section{Conclusion}\label{sec:conclusion}
We have investigated the quadrupole moments of Dy $5d$ and Fe $4p$
electrons in \DFBO\ which has the chiral helix structure of Dy and Fe
ions on the screw axes. 
We have performed resonant x-ray diffraction with circularly polarized
x-ray beam at Dy $L_3$ and Fe $K$ absorption edges by swithching the
sign of the x-ray helicity. 
The integrated intensity of forbidden reflections 004 and 005 has been
observed as a function of azimuth angle $\Psi$ and temperature.

The integrated intensity of the diffracted beam is theoretically
interpreted well:
(i)  the periodicity of azimuth angle scan is 120 degree, which agrees
with crystal symmetry,  
(ii) the helicity of circular polarization changes the intensity
according to the crystal handedness, 
(iii) the phase shift $\phi$ in azimuthal scan is unchanged for both two 
helicity states $(-)$ and $(+)$ of the x-ray beam at Dy $L_3$ and Fe
$K$ edges, 
(iv) the intensity at Dy $L_3$ edge is described by two components of
the Dy $5d$ quadrupole,
(v) on the other hand, that at Fe $K$ edge is described by totally six
components of the Fe $4p$ quadrupoles for three Fe ions at the special
position $3a$ and six Fe ions at the general position $6c$. 

From the results (i) and (ii), we find that the crystal which we
observed belongs to the left-handed space group $P3_221$, which
is in accord with that we determined with soft x-ray diffraction at Dy
$M_5$ absorption edge.~\cite{Usui2014ooq}
The result (iii) evidences that the resonant processes at both Dy $L_3$
and Fe $K$ edges is well described within the scheme of $E1E1$ process,
and that the birefringence phenomenon, or higher-order transition
processes like $E1E2$ is unnecessary to be considered for the analysis
of the experimental data. 
The discrepancy of the phase shift $\phi$ between reflections 004 and
005 observed at Fe $K$ edge is theoretically predicted by the admixture
of Fe $4p$ quadrupole moments at two sites.

From the result (iv), we have determined the ratio of two components 
of the Dy $5d$ moment at $T=100$ K 
as $r=\left | {\left<T_{+1}^2\right >''} / {\left <T_{+2}^2 \right >' } \right |=0.67\pm 0.02$,
which is smaller than $r=1.20$ for the Dy $4f$ moment.~\cite{Usui2014ooq}
The temperature dependence of the diffracted intensity shows 
a rather monotonous increase on cooling after the jump just 
below the $T_S$, while that of reflection 001
observed at Dy $M_5$ edge shows a steep increase towards lower
temperatures.
These results indicate that the Dy $4f$ moment is less coupled with
ligand oxygen atoms than the Dy $5d$ moment.

\begin{acknowledgments}
It is a pleasure to acknowledge important discussions with 
M. Takahashi.
This work was  supported by Grants-in-Aid for Scientific Research
 Nos. 24244058, and 25247054,  MEXT, Japan.
Resonant x-ray diffraction  experiments were performed at beam lines
 29XUL and 17SU in SPring-8 
(RIKEN Proposal Nos. 20120019, and 20130066) and beam line 3A in PF.
\end{acknowledgments}

\appendix
\section{Calculation for the components of Dy $5d$ moment using the
least squares method}\label{appendix:solve} 

The experimental x-ray polarization was not purely defined 
as $P_1=P_3=0$, $P_2=\pm1$.
We use the values of $P_2$ and $P_3$ summarized in Table~\ref{tab:polarization} for the analysis.
The value of $P_1$ is negligible.  There are two experimental parameters,
$p=\frac{9}{4} \left < T_{+2}^2  \right > '^2$, 
and $q=\frac{9}{4}\left <T_{+2}^2 \right >'\left <T_{+1}^2 \right >''$. 
The difference integrated intensity $\Delta I$ between two helicity states is 
\begin{equation}
\begin{aligned}
\Delta I =& \Delta I_0 + \Delta I_1 \cos 3( \lambda \Psi +\phi),  \label{isa} \\  
\Delta I_0 =& I_0 ^- - I_0^+ \\
=&  \left \{ \frac{1}{2}   (P_3^- -P_3^+) \cos ^2 \theta + \lambda (P_2^- - P_2^+)  \sin \theta   \right \} \\ 
& \times     (1+\sin ^2 \theta) \, p,       \\
\Delta I _1 = & I_1^- - I_1^+ \\ 
 = & \left \{ 2(P_3^- - P_3^+)\sin \theta - \lambda (P_2^- -P_2^+)  \cos ^2 \theta \right \} \\
& \times  \cos \theta \, q.    
\end{aligned}
\end{equation}
Here the suffixes `$+$' and `$-$' represent the positive, $P_2>0$, and
negative, $P_2<0$, states, respectively. 
The sum of the squares of the difference between the experimental data
and the theoretical functions for reflections 004 and 005 is 
\begin{equation}
\Delta = ({}^{\text{exp}}\Delta I_{004} -{}^{\text{cal}}\Delta I_{004})^2 
 +({}^{\text{exp}}\Delta I_{005} -{}^{\text{cal}}\Delta I_{004} )^2.
\end{equation}
Here the prefixes `exp' and `cal' represent the experimental data set
selected for the fit, and Eq.~\ref{isa} including parameters $p$ and
$q$, respectively.  
The least squares method is carried out to make the sum $\Delta$
minimum.  
This is a simple linear-least-squares calculation, and the values of $p$
and $q$ are directly obtained. 
Finally, we find that $p=135.7\pm 2.14$ and $q=90.8\pm 2.46$, and that 
the components of the Dy $5d$ quadrupole are 
$\left |\left <T_{+1}^2\right >\right |''=5.20\pm 0.15$, 
$\left | \left <T_{+2}^2 \right >\right |' = 7.77\pm 0.061$, 
 and
$r =\left | {\left<T_{+1}^2\right >''} / {\left <T_{+2}^2 \right >' }
\right | =0.67\pm 0.02$.
Note that the sign of each component is undetermined because the sign of
$I_1$  depends on the arbitrariness of  $\phi$ by 60 degrees.
 
\bibliography{dyfe3}

\end{document}